# New Use Cases for Snort: Cloud and Mobile Environments

Mayank Kumar, Emre Erturk

## Abstract

First, this case study explores an Intrusion Detection System package called Snort (provided by Cisco Systems) in a cloud environment. Snort is an open source and highly scalable signature-based intrusion detection system. Here, Snort is deployed on Ubuntu Server 16.0.4 running on a virtual machine within a Microsoft Azure cloud system. This paper provides details on installing Snort on the virtual machine and configuring it for intrusion detection. The architecture here is based on a VM integrated IDS on Azure and demonstrates how a VM instance in the cloud can be secured through an IDS. Firewalls may be considered the first line of defense but they fail to secure systems from inside attacks. Next, two other areas (where Snort is less widely used) are briefly explored, namely library systems and mobile devices. Finally, this paper makes further recommendations on how a cloud network can be secured by distributed placement of the IDS and on each VM instances.

## Introduction

Cloud computing has evolved into a new way of computing model providing resources and services on the internet. Three types of services provided over cloud are Software as a Service (SaaS), Platform as a Service (PaaS) and Infrastructure as a Service (IaaS) (Mell & Grance, 2011). Any attacks and vulnerabilities can affect the confidentiality, integrity and availability (a.k.a. CIA) of cloud resources. Obviously, the security of cloud services is one of the major challenges in going to the Cloud. Attacks may comprise of IP spoofing, DDoS, port scanning etc. Major cloud providers use firewall to prevent the outside attacks and is considered first line of defense. But firewall cannot detect insider attack from within the network and more complex forms of attack go un-noticed through firewalls. This calls for a more robust security system to be implemented and NIDS (Network intrusion detection system) performs exactly what is needed. In this case study, we use Snort (an open source NIDS) and explore the possibilities of securing the cloud system and propose some recommendations to secure cloud services.

Snort (a product from Cisco) is a signature-based intrusion detection system which allows monitoring of network traffic. It analyzes network traffic for any type of intrusion and generates alerts. It is an open source NIDS available under GPL. It is available for Microsoft Windows and Mac OS, as well as Linux based operating systems.

**Snort Architecture**

As described also in the Snort manual (available on https://www.snort.org/), Snort comprises of mainly five components working together to monitor and analyze all network traffic and look for any signs of intrusions and generate alert. As shown in figure 1 below, the five major components are as follows:

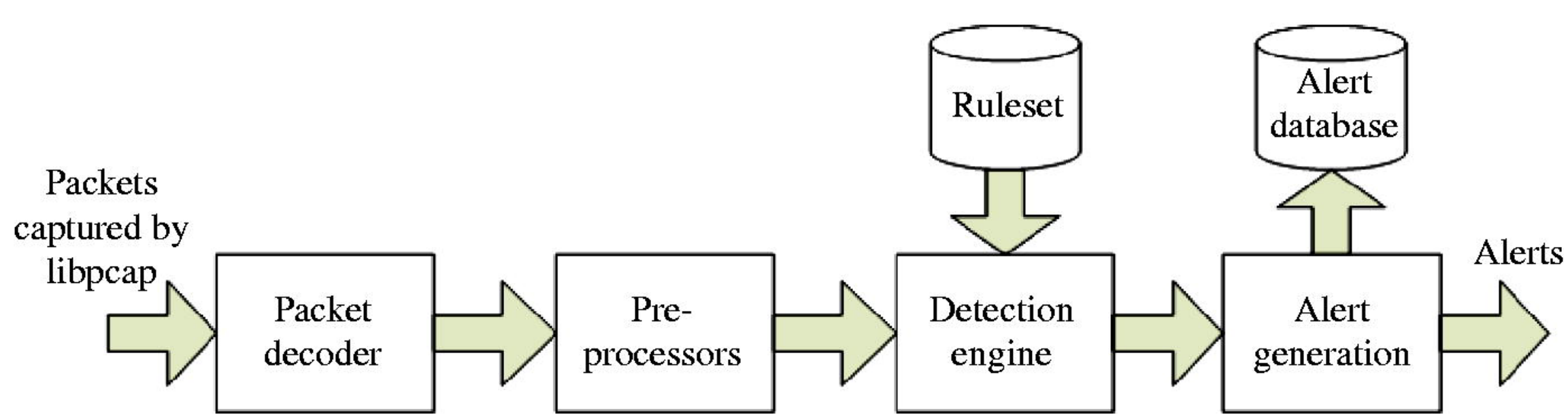

**Figure 1**: Snort architecture (Source: Snort Manual)

*Packet Decoder*: Captures packets from the network traffic and sets them up for preprocessor or for the detection engine.

*Preprocessors*: Processes the captured packets against certain plugins. These plugins check for known type behavior or anomalies. Preprocessors are indispensable part of any IDS to prepare data packets to be checked by detection engine against rules in the detection engine as intruders may modify those packets to escape any detection.

*Detection Engine*: Once the data is handled by the preprocessors, it is then passed on to detection engine. The detection engine is the most critical component of the signature-based IDS in Snort. It matches the data packets with the set of rules for any intrusion signatures contained in the data packets. If the rules match the data packets then it is passed on to the alert processor. It may take different amount of time for responding to several types of packets irrespective of the computing system it is running on.

*Logging and Alerting System*: Generation of alerts and logging is handled by this system. All the alerts and logs are kept in simple plain text files or tcp-dump style files.

*Output Module*: Output module helps save the logs generated by logging and alerting system in diverse ways like in simple plain text log files, logging to database like MySQL or Oracle or generating XML depending on the configuration set into Snort configuration file.

**IDS Architecture for Cloud**

In Figure 2 (Roschke, Feng, & Meinel, 2009) proposed a virtual machine integrated IDS architecture. It consists mainly of two components, IDS management unit and IDS sensor. Event gatherer built in IDS management unit collects anomalous behavior detected by IDS sensor and stores into event database. Analysis component then studies these logged events in the database according to the configuration. The IDS Remote Controller can communicate with IDS-VMs and IDS sensors, and hence manages the IDS-VMs. This approach of IDS deployment helps secure the virtual machines in the cloud and the services running on it.

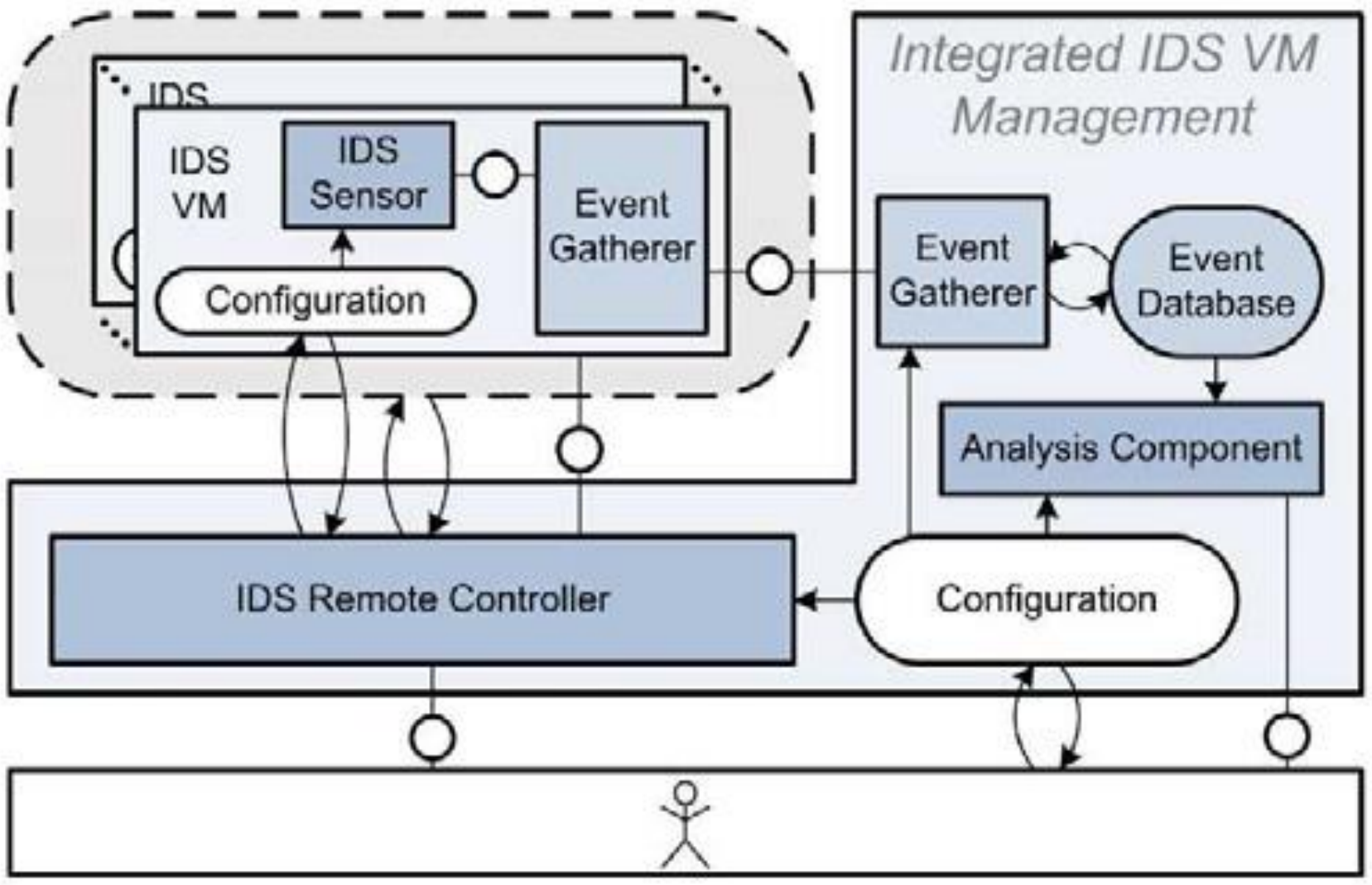

**Figure 2**: VM integrated IDS architecture (Source: Modi et al., 2012)

**Implementation**

In this experimental setup, we have chosen Microsoft Azure as our cloud platform and created a Linux Ubuntu Server 16.0.4 LTS virtual machine using a 30-day free trial, which allows the user to choose from various plans and features to incorporate into the virtual machine depending upon the computing resources required. Using Linux for Snort is a logical choice among practitioners. For example, Karim, Vien, Le, and Mapp (2017) found that using Linux to run Snort provides an improved performance of up to 10% over other operating systems. Anyhow, the setup below takes only 10-15 minutes, and is done on a well-defined user interface and guidance system, which help create the virtual machine relatively easily.

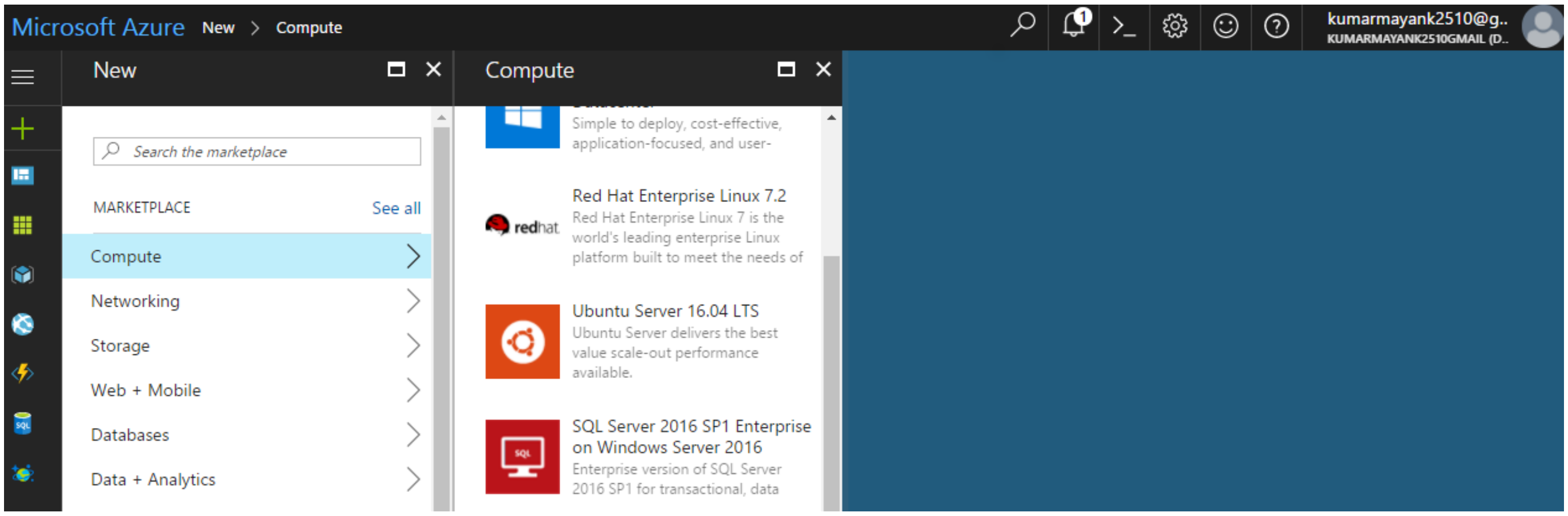

**Figure 3**: Virtual machine setup on Microsoft Azure.

Once the virtual machine is up and running, we start the installation of Snort. But before we can install Snort, we need to install some pre-requisites.

<u>Snort has 4 essential pre-requisites:</u>

| | |
|---|---|
| pcap (libpcap-dev) | available from the Ubuntu repository |
| PCRE (libpcre3-dev) | available from the Ubuntu repository |
| Libdnet (libdumbnet-dev) | available from the Ubuntu repository |
| DAQ | (http://www.snort.org/downloads/) compiled from source |

Firstly, we need to install all tools required for building software.

```
sudo apt-get install -y build-essential
```

Once build-essential is installed, we now install all the prerequisite packages which are available at Ubuntu repositories.

```
sudo apt-get install -y libpcap-dev libpcre3-dev libdumbnet-dev
```

The Snort DAQ (Data AcQuisition library) needs few pre-requisites that need to be installed:

```
sudo apt-get install -y bison flex
```

Install the latest version of DAQ using following set of commands. The PCAP DAQ module is the default module, used for getting packets into Snort from a file or an interface.

```
wget https://snort.org/downloads/snort/daq-2.0.6.tar.gz
tar -xvzf daq-2.0.6.tar.gz
cd daq-2.0.6
./configure
make
sudo make install
```

*Installing Snort:*

Four optional libraries that improves functionality are: liblzma-dev three of which provide decompression of swf files (adobe flash), openssl, and libssl-dev which both provide SHA and MD5 file signatures:

```
sudo apt-get install -y zlib1g-dev liblzma-dev openssl libssl-dev
```

After all pre-requisites are installed, now we are ready to download the Snort source tarball, compile, and install. The --enable-Sourcefire option gives Packet Performance Monitoring (PPM), which lets us do performance monitoring for rules and pre-processors, and builds Snort the same way that the Snort team does:

```
mkdir ~/snort_src
cd ~/snort_src
wget https://snort.org/downloads/snort/snort-2.9.9.0.tar.gz
tar -xvzf snort-2.9.9.0.tar.gz cd snort-2.9.9.0
 ./configure --enable-sourcefire
make
sudo make install
```

Update the shared libraries by running the following command (skipping this step will result in error while running Snort):

```
sudo ldconfig
```

Place a symlink to the Snort binary in /usr/sbin:

```
sudo ln -s /usr/local/bin/snort  /usr/sbin/snort
```

Test Snort by running the binary as a regular user, passing it the -V flag (which tells Snort to verify itself and any configuration files passed to it).

```
snort -V
```

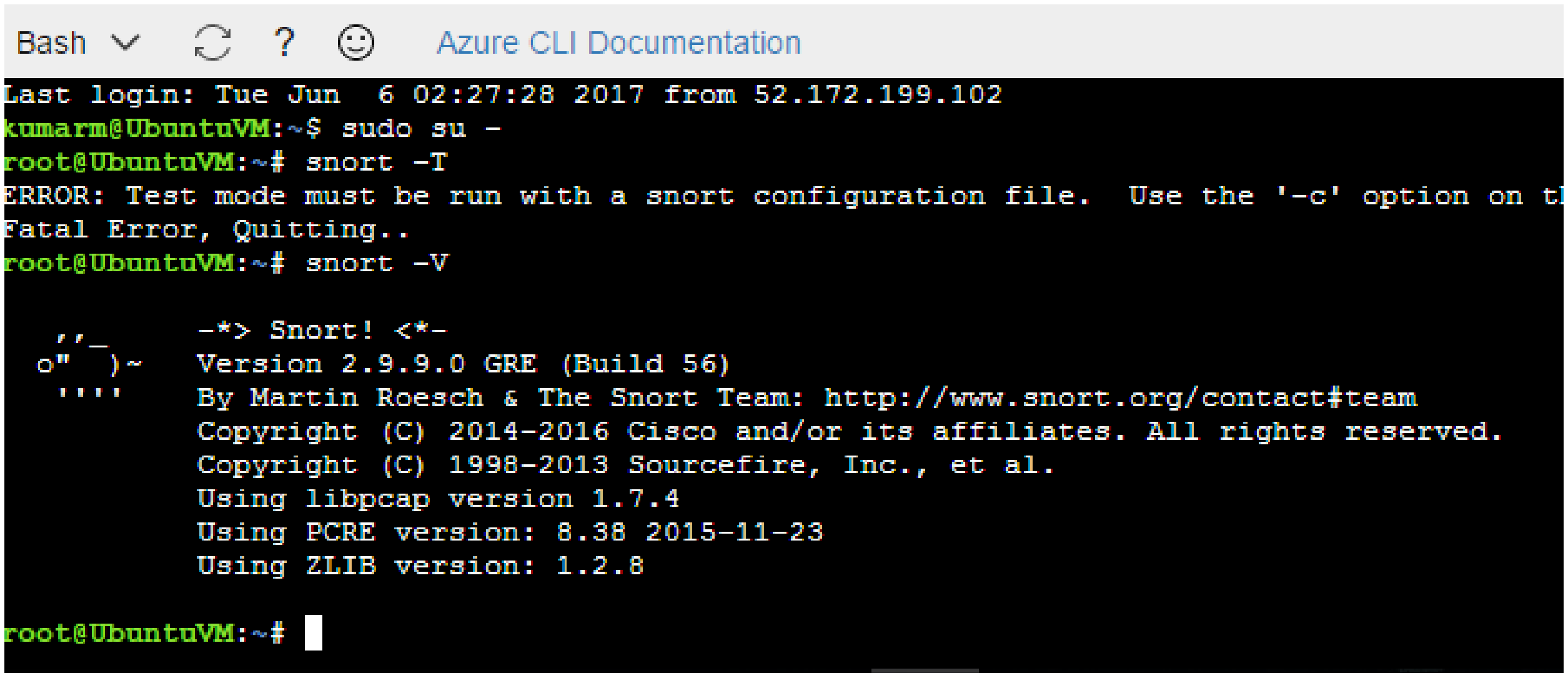

**Figure 4**: Snort installation validation.

*Configuring Snort:*

Create few directories and files as shown below for Snort and setup their permission.

```
# Create the Snort directories:
sudo mkdir /etc/snort
sudo mkdir /etc/snort/rules
sudo mkdir /etc/snort/rules/iplists
sudo mkdir /etc/snort/preproc_rules
sudo mkdir /usr/local/lib/snort_dynamicrules
sudo mkdir /etc/snort/so_rules
# Create files that store rules and ip lists
sudo touch /etc/snort/rules/iplists/black_list.rules
sudo touch /etc/snort/rules/iplists/white_list.rules
sudo touch /etc/snort/rules/local.rules
sudo touch /etc/snort/sid-msg.map
# Create our logging directories:
sudo mkdir /var/log/snort
sudo mkdir /var/log/snort/archived_logs
# Adjust permissions:
sudo chmod -R 5775 /etc/snort
sudo chmod -R 5775 /var/log/snort
sudo chmod -R 5775 /var/log/snort/archived_logs
```

```
sudo chmod -R 5775 /etc/snort/so_rules
sudo chmod -R 5775 /usr/local/lib/snort_dynamicrules
```

To copy the configuration files and the dynamic preprocessors, run the following commands:

```
cd ~/snort_src/snort-2.9.9.0/etc/
sudo cp *.conf* /etc/snort
sudo cp *.map /etc/snort
sudo cp *.dtd /etc/snort
cd ~/snort_src/snort-2.9.9.0/src/dynamicpreprocessors/build/usr/local/lib/snort_dynamicpreprocessor/
sudo cp * /usr/local/lib/snort_dynamicpreprocessor/
```

Edit Snort's main configuration file, snort.conf. To run Snort in NIDS mode, this file needs to be passed as an argument.

Next, open /etc/snort/snort.conf file in any editor:

```
nano /etc/snort/snort.conf
```

Provide the machine's ip address and define paths as shown below:

```
# Setup the network addresses you are protecting
ipvar HOME_NET 10.0.0.4/24

# Set up the external network addresses. Leave as "any" in most situations
ipvar EXTERNAL_NET any

var RULE_PATH /etc/snort/rules
var SO_RULE_PATH /etc/snort/so_rules
var PREPROC_RULE_PATH /etc/snort/preproc_rules
var WHITE_LIST_PATH  /etc/snort/rules
var BLACK_LIST_PATH /etc/snort/rules
include $RULE_PATH/local.rules
```

Save and close the file. Next, using below command validate the configuration file:

```
Sudo  /usr/local/bin/snort -A console -q -u snort -g snort -c /etc/snort/snort.conf -i eth0
```

```
Bash ⌄   ?   Azure CLI Documentation
            Preprocessor Object: SF_SSLPP  Version 1.1  <Build 4>
            Preprocessor Object: SF_DCERPC2  Version 1.0  <Build 3>
            Preprocessor Object: SF_SSH  Version 1.1  <Build 3>
            Preprocessor Object: SF_DNS  Version 1.1  <Build 4>
            Preprocessor Object: SF_MODBUS  Version 1.1  <Build 1>
            Preprocessor Object: SF_IMAP  Version 1.0  <Build 1>
            Preprocessor Object: SF_SIP  Version 1.1  <Build 1>
            Preprocessor Object: SF_SDF  Version 1.1  <Build 1>
            Preprocessor Object: SF_REPUTATION  Version 1.1  <Build 1>
            Preprocessor Object: SF_SMTP  Version 1.1  <Build 9>
            Preprocessor Object: SF_GTP  Version 1.1  <Build 1>
            Preprocessor Object: SF_FTPTELNET  Version 1.2  <Build 13>
            Preprocessor Object: SF_DNP3  Version 1.1  <Build 1>

Snort successfully validated the configuration!
Snort exiting
root@UbuntuVM:/pulledpork/pulledpork-master#
```

**Figure 5**: Snort configuration file validation.

*Testing Snort*

Now Snort can be tested but before that few rules need to be created to test local data packets on the network.

Lastly, create some rules to test Snort.

First, edit the local.rules file:

```
nano  /etc/snort/rules/local.rules
```

Add the following lines:

```
Bash ⌄   ?   Azure CLI Documentation
  GNU nano 2.5.3                              File: /etc/snort/rules/local.rules

# $Id: local.rules,v 1.11 2004/07/23 20:15:44 bmc Exp $
# ----------------
# LOCAL RULES
# ----------------
# This file intentionally does not come with signatures.  Put your local
# additions here.
alert tcp any any -> any any (msg:"FTP connection attempt"; sid:1000001; rev:1;)

alert icmp any any -> any any (msg:"ICMP connection attempt"; sid:1000002; rev:1;)

alert tcp any any -> any any (msg:"TELNET connection attempt"; sid:1000003; rev:1;)
```

**Figure 6**: User created sample rules in local.rules file.

Save and close the file.

The above rules will generate alerts when someone tries to Ping, FTP or Telnet to the server.

Now, run Snort in NIDS mode and send alert output to the console:

```
snort -A console -q -c /etc/snort/snort.conf
```

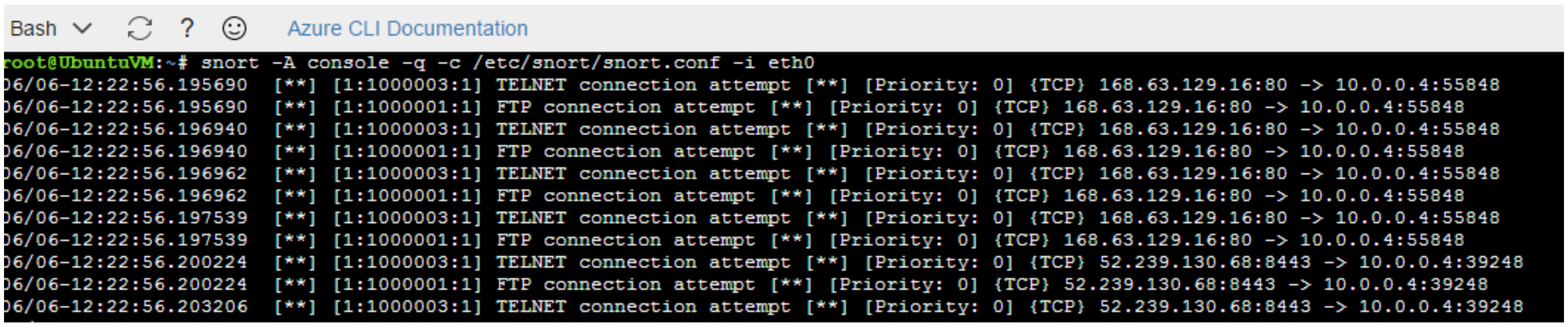

**Figure 7**: Alerts generated by Snort in NIDS using local rules created.

*Snort in sniffer mode:*

```
Snort -vde
```

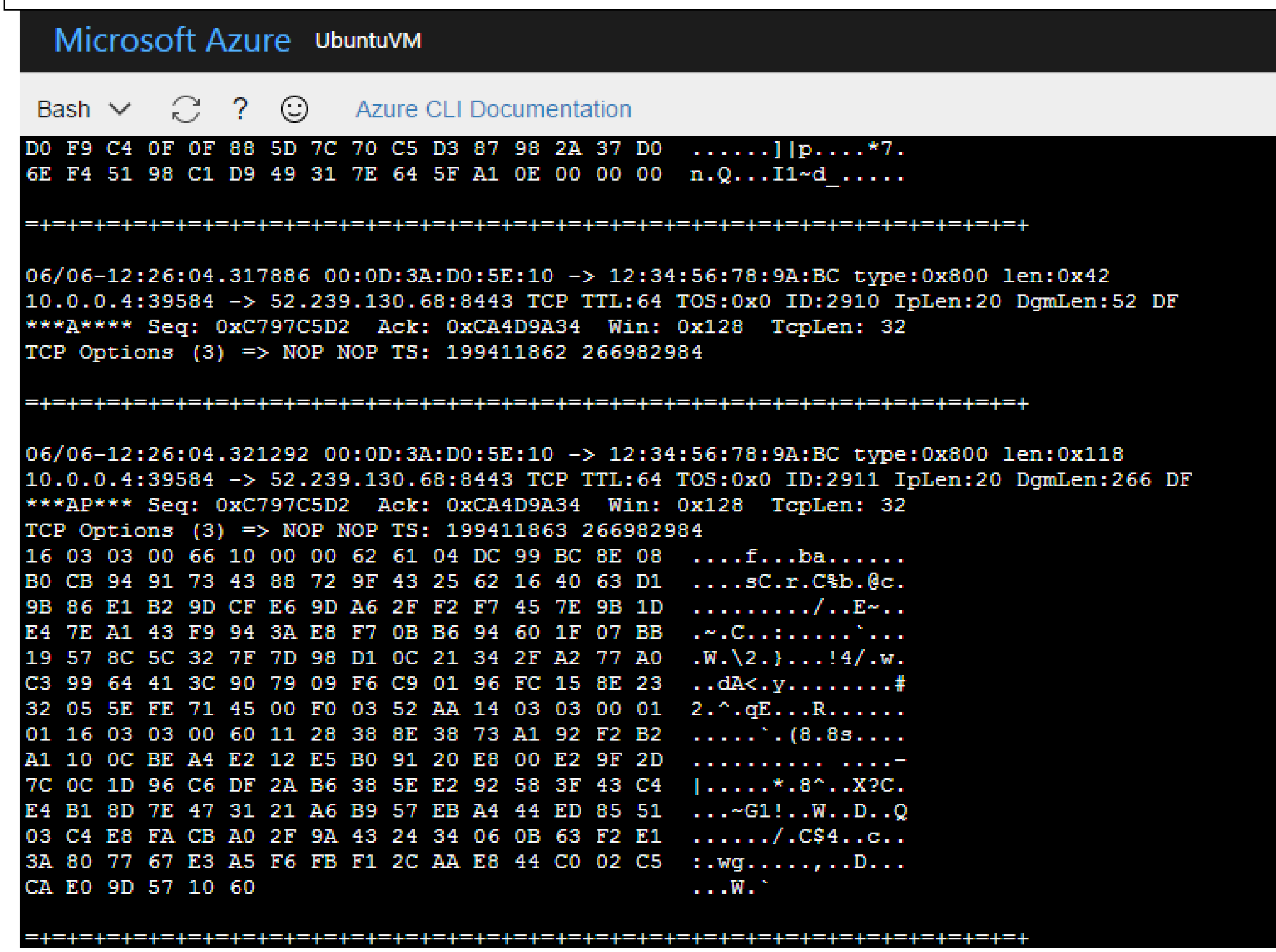

**Figure 8**: Snort in sniffer mode.

## Discussion and Use Cases

Snort is a widely used open source NIDS, supporting various operating system environments. With many plugins supported by Snort, its functionality and efficiency can be adapted to new and different deployment environments. It can act as a host-based intrusion detection system when installed on virtual machines in the Cloud or as a network intrusion detection system when installed on network interface cards or routers. One of the major advantages of Snort is its ability to be configured as per the needs of any organization, and creation of own rule sets which can be used by advanced users to implement soft computing techniques and generate their own rule sets to be used with Snort. It is very scalable, has a low computational cost, fewer false positives and low false negatives. During the implementation, Snort could detect traffic through the network interface passing through the virtual machine in that Azure Cloud. Additionally, it was able to distinctly identify several types of data packets such as TCP and TELNET on the network and generated custom message alerts as defined in the local rules.

Snort can be used by a variety of organizations and businesses. However, an area of oversight (and a new use case) is using Snort as part of the security of library information systems. A typical library information system may include an EZproxy component for authenticating and providing online library resources to off-campus users (Erturk & Iles, 2015). Snort, for example, can be used to analyze EZproxy packets when there are unusual issues (Feinstein, 2006). This is a technical area that requires more documentation, and can be studied in the future with the most recent versions of Snort.

Snort has been used very little on mobile devices, as it may have been considered to require more powerful resources, and probably not even many Android users were interested in it, functionally or as part of a technical security hobby. On the other hand, the open source nature of the Android operating system and the presence of certain threats allow greater innovation for developers and users (Erturk, 2012). In this case, the novelty would be to run Snort or similar tools in a mobile environment. An early example of porting Snort as an Android client app has been Swinedroid, which can monitor and help protect a server remotely (Budington, 2011). More recent examples (not based on Snort) of IDS tools available in the Google app market are PortDroid and LogDog (https://play.google.com). Finally, a sensitive area particularly for mobile applications is location based services. According to a recent work by Ho Kang and Park (2017), one way to help secure these services is to utilize Snort to protect the application server and client devices from intrusion by setting up sophisticated rules to scan these packets.

## Conclusion

Although installing Snort on a virtual machine in the Cloud provides protection to that machine from malicious activity, it cannot detect intrusion coming from outside of a network. As a result, a more distributed form of IDS placement is required for a cloud network (Modi et al., 2012). The placement of IDS on a central server can help in detecting both internal as well as external intrusions. Snort can also be placed on a reverse proxy, i.e. a server that acts as an intermediary between a completely backend server and the clients (Bakhdlaghi, 2017). Also, placing the IDS on each VM instance provides more robust security to a cloud computing system. Snort can also be used together with a relational database such as MySQL or Oracle to log the alerts for further analysis by another special plugin. Snort provides constant updates to the rule sets to include new threat signatures. These can be downloaded manually and updated, or the Pulled Pork plugin can be used to maintain and update the rule sets. Snort is a light weight and highly configurable IDS (with many plugins that enhance its functionality). As the popularity of cloud services grows, an IDS will play an important role in addressing the ongoing security requirements.

**References**

Bakhdlaghi, Y. (2017). Snort and SSL/TLS Inspection. Retrieved from https://www.sans.org/reading-room/whitepapers/detection/snort-ssl-tls-inspection-37735

Budington, W. (2011). Swinedroid – the new Snort Monitoring tool for Android. Retrieved from https://www.inputoutput.io/swinedroid-the-new-snort-monitoring-tool-for-android/

Erturk, E. (2012). Two Trends in Mobile Malware: Financial Motives and Transitioning from Static to Dynamic Analysis. Infonomics Society.

Erturk, E., & Iles, R. (2015). Case study on cloud based library software as a service: Evaluating EZproxy. Journal of Emerging Trends in Computing and Information Sciences, 6(10), 545-549. Retrieved from https://arxiv.org/ftp/arxiv/papers/1511/1511.07578.pdf

Feinstein, M. (2006, September 26). IPTABLES INVALID state. Retrieved from https://lists.gt.net/iptables/user/62492

Ho Kang, B., & Park W. (2017). An Enhancement of Optimized Detection Rule of Security Monitoring and Control for Detection of Cyberthreat in Location-Based Mobile System. *Mobile Information Systems.*

Jethva, Hitesh. (2017). How to Install Snort NIDS on Ubuntu Linux. https://komunity.komand.com/learn/featured/how-to-install-snort-nids-on-ubuntu-linux/

Karim, I., Vien, Q.-T., Le, T., & Mapp, G. (2017). A Comparative Experimental Design and Performance Analysis of Snort-Based Intrusion Detection System in Practical Computer Networks. *Computers, 6*(1), 6.

Mell, P. and T. Grance. (2011).The NIST definition of cloud computing. *National Institute of Standards and Technology*. NIST Special Publication 800-145.

Modi, Chirag. N et al.(2012). Integrating Signature Apriori based Network Intrusion Detection System (NIDS) in Cloud Computing. *2nd International Conference on Communication, Computing & Security* (ICCCS-2012).

Roschke, S., Feng, C., & Meinel, C. (2009). An Extensible and Virtualization Compatible IDS Management Architecture. *International Conference on Information Assurance and Security*. pp. 130–134.

Snort Manual. (n.d). Snort Homepage. https://www.Snort.org/.